\documentclass[a4paper,12pt]{article}
\newif\ifpdf            
\ifx\pdfoutput\undefined \pdffalse \else \pdftrue \fi
\ifpdf 
\usepackage[pdftex]{graphicx}
\pdfcompresslevel9 \DeclareGraphicsExtensions{.jpg,.pdf,.png,.mps}
\else 
\usepackage[dvips]{graphicx}
\DeclareGraphicsExtensions{.eps,.ps} \fi
\usepackage{bm}

\newcommand{\beq}{\begin{equation}}
\newcommand{\eeq}{\end{equation}}

\renewcommand{\ll}{\ensuremath{\underline{\underline{\lambda}}}}

\begin{document}

\title{A Tube Model of Rubber Elasticity}
\author{S.~Kutter and
E.M.~Terentjev\footnote{Electronic mail: {\it emt1000@cam.ac.uk}} \\
Cavendish Laboratory, University of Cambridge \\ Madingley Road,
Cambridge CB3 0HE, U.K. }

\date{\today}
\maketitle

\begin{abstract}
\noindent Polymer entanglements lead to complicated topological
constraints and interactions between neighbouring chains in a
dense solution or melt. Entanglements can be treated in a mean
field approach, within the famous reptation model, since they
effectively confine each individual chain in a tube-like geometry.
In polymer networks, due to crosslinks preventing the reptation
constraint release, entanglements acquire a different topological
meaning and have a much stronger effect on the resulting
mechanical response. We apply the classical ideas of reptation
dynamics to calculate the effective rubber-elastic free energy of
an entangled rubbery network. We then compare the results with
other theoretical approaches and establish a particularly close
mapping with the hoop-model, with equally good description of
experimental data. The present consistent reptation theory allows
further development of dynamic theory of stress
relaxation.
\end{abstract}
 \vspace{0.2cm}
        {62.20.Dc} \ {Elasticity, elastic constants}

\newpage

\section{Introduction}

Rubbery polymer networks are highly complex disordered systems.
The simplest theoretical models consider them as being
made of ``phantom chains'', where
each polymer is modelled by a three-dimensional random walk in
space. To form a corresponding phantom network, the chains are
crosslinked to each other at their end points, but do not interact
otherwise, in particular, they are able to fluctuate freely
between crosslinks. This has the unphysical consequence that the
strands can pass through each other. If one tries to avoid this
assumption, the theory is confronted with the intractable
complexity of entanglements and their topological constraints. The
mean field treatment of entangled polymer systems is a now
classical reptation theory (de Gennes 1979; Doi \& Edwards 1986),
which had a spectacular success in describing a large variety of
different physical effects in melts and semi-dilute solutions.
However, the parallel description of crosslinked rubbery networks
has been much less successful. First of all, one has to appreciate
a significant difference in the entanglement topology: in a
polymer melt the confining chain has to be long enough to form a
topological knot around a chosen polymer; even then the constraint
is only dynamical and can be released by a reptation diffusion
along the chain path. In a crosslinked network, any loop around a
chosen strand becomes an entanglement, which could be mobile but
cannot be released altogether. A number of other complexities
arise from possible nematic interactions between rigid chain
segments (Abramchuk \textit{et al.} 1989; Bladon \& Warner 1993)
and from the coupling between imposed deformations and chain
anisotropy known as the stress-optical effects (Jarry \& Monnerie
1979; Deloche \& Samulski 1981, Doi \textit{et al.} 1989).

An early model of elastic response of entangled rubbers was
developed by Edwards (1977): in tradition with the melt
theory, it assumed that the presence of neighbouring strands in a
dense network effectively confines a particular polymer strand to
a tube, whose axis defines the primitive path, see Fig.~\ref{pic}.
Within this tube, the polymer is free to explore all possible
configurations, performing random excursions, parallel and
perpendicular to the axis of the tube. One can show that on
deformation the length of the primitive path increases. Since the
arc length of the polymer is constant, the amount of chain
available for perpendicular excursions is reduced, leading to a
reduction in entropy and hence to an increase in free energy.
However, the particular calculation in Edwards (1977) has a
number of shortcomings; perhaps the main limitation is that one
only looks at the entropy reduction associated with the overall
change of primitive path contour length on deformation, ignoring
the essential mechanisms of local reptation and segment
re-distribution between different tube segments.

Gaylord and Douglas (GD) (1990) developed a simple
``localisation model'' for rubbers, based on scaling arguments.
Each strand segment is thought to be placed in a hard tube of a
square cross-section. Assuming that the deformation affinely
changes the dimensions of the tube, one can calculate the change
in free energy of a network of chains, each confined in such a
tube. Ball, Doi, Edwards and Warner (BDEW) (1981)
chose a completely different approach by introducing the
``slip-link model'': the effect of entanglements is not considered
to be a permanent one, changing the environment of a single
polymer strand as in the two models presented above, but rather
leading to a local mobile confinement site, a link between two
interwound strands, which is able to ``slip'' up and down along
both strands. In a further development of this idea, Higgs and Ball (HB)
(1989) adopted a similar approach, which is however
mathematically much simpler: the entanglements localise certain
short segments of a particular strand to a small volume. One can
model this effect by describing a network strand as a free
Gaussian random walk, which is, however, forced to pass through a
certain number of hoops, which are fixed in space. We shall find
that the results of HB are very close to ours, in spite of a
number of significant differences in the physical model. This
mapping gives confidence in the final expression for the
rubber-elastic free energy and the role of network entanglements
in it.

In addition to the mentioned above, one can find a variety of
other theoretical models, some of which are discussed in the
review article (Edwards \& Vilgis 1988). For instance, the
constrained junction fluctuation model (Flory \& Erman 1982),
which assumes that the entanglements primarily affect the
fluctuation of the junction points. An overview of different
models can be found for example in the reviews by Heinrich
\textit{et al.} (1995) and Han \textit{et al.} (1999). All
theoretical models, describing the macroscopic equilibrium elastic
response of densely entangled rubbery networks, have a common
purpose -- to develop a physically consistent
description depicting what one accepts as a correct coarse-grained
molecular behaviour -- but also accounting for a number of
experimental results showing substantial deviations in
stress-strain response from the ideal phantom-network result
$\sigma = \mu (\lambda - 1/\lambda^2)$ (Higgs \& Gaylord 1990). Up
to now, no theory succeeded on both of these fronts.

In our current work, we develop a consistent implementation of the
classical tube model to take into account the entanglement effects
in crosslinked polymer networks within the same framework as in
the polymer melt dynamics. We particularly focus on the entropy of
internal reptation motion and resulting re-distribution of chain
segments along the tube in a deformed state.
The reptation tube model of entangled network strands provides a more
accurate description in the sense that it keeps track of the
allocation of chain segment excursions in the tubes.
In this way, our
model closes an important gap among the existing models with some
unexpected consequences: the model, although based on existing ideas, analyses
rubber elasticity in a new way. The calculations, however, reveal that
our results are very similar to the ones of the HB model, which approaches
rubber elasticity in a different way.
After brief recollection of the principles of network theory on the
example of the ideal phantom network in the following section, section 3
introduces the model and its properties in some detail, and
outlines the derivation of the full expression for rubber-elastic
free energy. Section~4 explores the properties of the full
expression, examining different limiting cases, as well as its
linear-response limit. We conclude by comparing the concepts and
the results of this work with previous theories.

\section{Classical phantom chain network}

Before considering densely entangled rubber, we briefly review the
well-known results of the phantom chain network theory, which
provides the basics to most other theoretical models.

Assuming that a single polymer performs a free random walk in
three dimensions, one finds that the end-to-end distance ${\bm
R}_0$ obeys  a Gaussian distribution in the long chain limit. This
result goes back far in history: one can review its derivation and
consequences in the classical text on this subject (Doi \& Edwards
1986). The distribution of $\bm{R}_0$ is given by
\begin{equation}
    P_0({\bm R}_0) = \left(\frac{3}{2\pi N
    b^2}\right)^{3/2}\exp\left(-\frac{3}{2 N b^2}{\bm R}_0^2\right)
 \ \ \propto \ Z(\bm{R}_0)=e^{-\beta F({ \bm R}_0)},
\label{gauss}
\end{equation}
where $b$ is the monomer step length and $N$ the number of steps
of the chain trajectory. The entropic free energy of such a random
walk, therefore, is given by the logarithm of the number of
conformations with the fixed $\bm{R}_0$ and has the form $\beta
F=-\ln P_0({\bm R}_0)=\frac{3}{2Nb^2}{\bm R}_0^2+\mathrm{const}$,
where $\beta=1/k_B T$ the inverse Boltzmann temperature. At
formation of the network, i.e. at crosslinking, each chain in the
polymer melt obeys the distribution (\ref{gauss}), which is then
permanently frozen in the network topology.

One then assumes that the network junction points deform affinely
with respect to their initial positions $\bm{R}_0$ following the
macroscopic deformation described by the tensor $\ll$; hence we
can write $\bm{R}=\ll \bm{R}_0$. Therefore, the deformation $\ll$
alters the free energy of each strand. The change of entropic free
energy per chain of the whole network can be calculated by the
usual quenched averaging:
\begin{equation}
   \beta F =-\langle\ln P(\bm{R})\rangle_{P(\bm{R}_0)}
   =\frac{1}{2}\mathrm{Tr}(\ll^{\mathsf{T}}\ll), \label{fgauss}
\end{equation}
where we have dropped an irrelevant constant, arising as the logarithm of the
normalisation in (\ref{gauss}). The overall elastic
energy density, in the first approximation, is simply
(\ref{fgauss}) multiplied by the number of elastically active
network strands in the system $n_{\rm ch}$ per unit volume, which
is proportional to the crosslinking density:
\begin{equation}
    F_{\rm elast}=
    \frac{1}{2}\mu \, \mathrm{Tr}\left(\ll^{\mathsf{T}}\ll\right),
    \quad\mbox{with}\quad
    \mu=n_{\rm ch}k_B T.\nonumber
\end{equation}

Three positive remarks have to be made in defence of this simple
model of rubber elasticity. First, consider the crosslinking
connecting the end points of different polymer strands. Since the
positions of end points of a single chain fluctuate strongly, the
junctions reduce these fluctuations and therefore alter the single
chain statistics. However, in spite of an apparent complexity,
this effect merely introduces a multiplicative
factor of the form $1-2/\phi$, where $\phi$ is the junction point
functionality (see for example the review by Flory (1976)).
Secondly, one
can assume that the deformation preserves the volume, since the
bulk (compression) modulus is by a factor of at least $10^4$
greater than the shear modulus, which is proportional to $\mu$;
this implies the constraint $\det\ll=1$. Thirdly, the quenched
average in equation (\ref{fgauss}) does not average over chains of
different arc lengths, but the fact that the result is independent
of arc length, generalises the result to apply for chains of
arbitrary length, or even for a polydisperse ensemble of chains.

As a result if one considers, for example, a uniaxial extension
$\lambda$ (by incompressibility, the two perpendicular directions
will experience an equal contraction $1/\sqrt{\lambda}$), the
elastic energy would take the form $F_{\rm elast} = \frac{1}{2}\mu
\left( \lambda^2 + 2/\lambda \right)$. This then allows us to
calculate the nominal stress, which is the force divided by the
area of the undeformed cross section:
 \begin{equation}
    \sigma_{\rm phantom}= \frac{\partial F}{\partial\lambda}=
        \mu\left(\lambda-\frac{1}{\lambda^2}\right)\label{sgauss}
 \end{equation}
Commonly, one draws the stress--strain curves in the Mooney-Rivlin
representation, which plots the reduced stress function
$f^*=\sigma/ (\lambda - 1/\lambda^2)$ against $1/\lambda$ (see
Fig.~\ref{mooney} in the discussion below). This representation
therefore indicates the degree of deviation from the simple
phantom chain network behaviour.

\section{Reptation theory of rubber elasticity}

\begin{figure}
\centering \resizebox{0.55\textwidth}{!}{\includegraphics{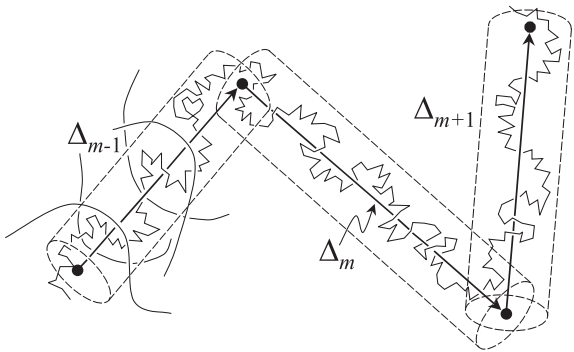}}
\caption{A polymer strand is surrounded by neighbouring chains,
which effectively confine the strand to a tube. The tube segment
$m$, with the span vector $\bm{\Delta}_m$ along its axis, contains
$s_m$ monomer steps, where $m$ runs from 1 to $M$.} \label{pic}
\end{figure}

Following the original ideas of Edwards (1977), we assume that
each network strand is limited in its lateral fluctuations by the
presence of neighbouring chains. Therefore each segment of a
polymer only explores configurations in a limited volume, which is
much smaller than the space occupied by an ideal random coil.
Hence, the whole strand fluctuates around a certain trajectory, a
mean path, which is called the primitive path in the reptation
theory (Edwards 1977). The most intuitive way to
visualise such a trajectory is to imagine that the chain is made
shorter and thus stretched between the fixed crosslinking points.
The taut portions of the chain will form the broken line of
straight segments between the points of entanglement, which
restrict the further tightening. This primitive path can be
considered as a random walk with an associated typical step
length, which is much bigger than the polymer step length, as
sketched in Fig.~\ref{pic}. The number of corresponding tube
segments $M$ is determined by the average number of entanglements
per chain (the situation with no entanglements corresponds to
$M=1$).

Effectively, the real polymer is confined by the neighbouring
chains to exercise its thermal motion only within a tube around
the primitive path. Note that all the chains are in constant
thermal motion, altering the local constraints they impose on each
other. Hence, the fixed tube is a gross simplification of the real
situation. However, one expects this to be an even better
approximation in rubber than in a corresponding melt (where the
success of reptation theory is undeniable), because the
restriction on chain reptation diffusion in a crosslinked network
eliminates the possibility of constraint release.

To handle the tube constraint mathematically, we traditionally
assume that the chain segments are subjected to a quadratic
potential, restricting their motion transversely to the primitive
path. Along one polymer strand consisting of $N$ monomers of
effective step length $b$, there are $M$ tube segments, each
containing $s_m,\ m=1, \dots, M$ monomer steps. We infer the
obvious condition
 \begin{equation}
    \sum_{m=1}^{M} s_m=N.\label{segmentsum}
 \end{equation}
In effect, one has two random walks: the topologically fixed
primitive path and the polymer chain restricted to move around it
-- both having the same end-to-end vector $\bm{R}_0$, between the
connected crosslinking points.

Each tube segment $m$ can be described by the span vector
$\bm{\Delta}_m$, joining the equilibrium positions of the strand
monomers at the two ends of each tube segments. The number of tube
segments $M$ (or, equivalently, the nodes of the primitive path)
is a free parameter of the theory, ultimately determined by the
length of each polymer strand and the entanglement density.

Since the primitive path is a topologically frozen characteristic
of each network strand, we shall assume that all primitive path
spans $\bm{\Delta}_m$ deform affinely with the macroscopic strain:
$\bm{\Delta}_m'=\ll \bm{\Delta}_m$. This is the central point in
the model: the rubber elastic response will arise due to the
change in the number of polymer configurations in a distorted
primitive path. To evaluate the number of conformations, we look
separately at chain excursions parallel and perpendicular to the
tube axis, along each span $\bm{\Delta}_m$. Effectively, this
amounts to introducing a new coordinate system for each tube
segment, with one preferred axis along $\bm{\Delta}_m$. In this
direction one recovers the behaviour of a free random walk, giving
rise to a one-dimensional Gaussian statistics in the long chain
limit. Note that only one third of the steps $s_m$ in this tube
segment would be involved in such parallel (longitudinal)
excursions. We therefore obtain for the number of parallel
excursions in a tube segment $m$:
 \begin{equation}
    W_m^{\parallel}\propto
    \frac{1}{\sqrt{s_m/3}}
    \exp\left(-\frac{1}{2b^2 (s_m/3)}\bm{\Delta}_m^2\right).
    \label{parallel}
\end{equation}

To determine the number of perpendicular (transverse) excursions,
one can introduce the Green's function for the perpendicular steps
made by the chain. In effect, we consider a two-dimensional random
walk, with a total number of steps $(2s_m/3)$, in a
centrosymmetric quadratic potential. For each of these two
perpendicular coordinates, the Green's function satisfies the
following modified diffusion equation (see e.g. Doi \& Edwards
1986):
 \begin{equation}
    \left(\frac{\partial}{\partial s}-
    \frac{b^2}{2}\frac{\partial^2}{\partial x_f^2}
    +\frac{q_0^2}{2}x_f^2\right)G(x_i,x_f;s)=\delta(x_f-x_i)\delta(s),
\label{greeneq}
 \end{equation}
where the $x_i$ and $x_f$ are the initial and final coordinates of
the random walk with respect to the tube axis  and $q_0$
determines the strength of the confining potential. This is an
essential parameter of many reptation theories, directly related
to the tube diameter $a$. Many discussions can be found in the
literature regarding, for instance, the possible dependence of $a$
on entanglement density and the possible change in $a$ on
deformation. However, we shall find that the potential strength
$q_0$ does not enter the final rubber-elastic energy, in an
approximation that we expect to hold for a majority of cases. Such
a universality resembles the situation with phantom networks,
where many chain parameters do not contribute to the final
results.

The equation (\ref{greeneq}) is very common in the physics of
polymers and its exact solution is known. However, we only need to
consider a particular limit $q_0 b s_m \gg 1$ of this solution,
which is the case of dense entanglements (resulting in a strong
confining potential) and/or of a large number $s_m$ of monomers
confined in the tube segment. Outside this limit, that is, when
the tube diameter is the same order as the arc length of the
confined chain, the whole concept of chain entanglements becomes
irrelevant. In the case of our present interest, that is, in the
strongly confined limit, the solution has a particularly simple
form, see (Edwards (1977),
 \begin{equation}
    G_m(x_i,x_f;s_m)\propto
    \exp\left(-\frac{q_0}{2 b}(x_i^2+x_f^2)-\frac{1}{6}q_0 b s_m\right).
 \end{equation}
Remembering that there are two coordinates describing the
transverse excursions, we obtain for the two-dimensional Green's
function of the tube segment $m$:
 \begin{equation}
    G_m(\bm{r}_i,\bm{r}_f;s_m)\propto
    \exp\left(-\frac{1}{3}q_0 b s_m\right)
    \exp\left(-\frac{q_0}{2 b}(\bm{r}_i^2+\bm{r}_f^2)\right),
    \label{2d-green}
 \end{equation}
where $\bm{r}_i$ and $\bm{r}_f$ are the initial and final
transverse two-dimensional coordinates.

The total number of transverse excursions is proportional to the
integrated Green's function:
 $$ W_m^{\perp}\propto\int d\bm{r}_i
\int d\bm{r}_f G_m(\bm{r}_i,\bm{r}_f;s_m).
 $$
Since the Green's function in the approximation (\ref{2d-green})
does not couple the initial or final coordinates to the number of
segments $s_m$, this integration will only produce a constant
normalisation factor which can be discarded.

Gathering the expressions for statistical weights of parallel and
perpendicular excursions, we obtain the total number of
configurations of a polymer segment consisting of $s_m$ monomers
in a tube segment of span $\bm{\Delta}_m$:
 \begin{equation}
    W_m=W_m^{\parallel} W_m^{\perp}\propto
    \frac{1}{\sqrt{s_m}}
    \exp\left(-\frac{3}{2b^2 s_m}\bm{\Delta}_m^2 -
    \frac{1}{3}q_0 bs_m\right).
 \end{equation}
Therefore, we find for the full number of configurations of the
whole strand by summing over all $M$ steps of the primitive path
and integrating over the remaining degrees of freedom, the number
of polymer segments confined between each node of the tube:
 \begin{equation}
 W= \int\limits_0^N ds_1 \cdots \int\limits_0^N ds_M
    \left(\prod_{m=1}^{M} W_m \right)
    \delta\left(\sum_{m=1}^{M} s_m-N\right),\label{no_config}
 \end{equation}
where the constraint (\ref{segmentsum}) on the polymer contour
length is implemented by the delta-function. The statistical
summation in (\ref{no_config}) takes into account the reptation
motion of the polymer between its two crosslinked ends, by which
the number of segments, $s_m$, constrained within each tube
segment can be changed and, thus, equilibrates for a given
conformation of primitive path.

Rewriting the delta-function as $\delta(x)=\frac{1}{2\pi} \int dk \
e^{{\rm i} kx}$, we proceed by finding the saddle points $s_m^*$
which make the exponent of the statistical sum (\ref{no_config})
stationary. It can be verified that the normalisation factors
$1/\sqrt{s_m}$ contribute only as a small correction to the saddle
points
\begin{equation}
s_m^* \approx \left( \frac{3\bm{\Delta}_m^2}{2b^2
    (\frac{1}{3}q_0b + {\rm i}k)}\right)^{1/2}.\label{saddle-sm}
 \end{equation}
The integral in (\ref{no_config}) is consequently approximated by
the steepest descent method. We repeat the same procedure for the
integration of the auxiliary variable $k$, responsible for the
conservation of the polymer arc length. The saddle point value
$k^*$, inserted back into (\ref{saddle-sm}), gives
the equilibrium number of polymer segments confined within a tube
segment with the span vector $\bm{\Delta}_m$:
\begin{equation}
    \overline{s_m}=\frac{N\Delta_m}{\sum_{m=1}^{M}\Delta_m} , \label{saddle-sm2}
\end{equation}
where $\Delta_m=|\bm{\Delta}_m|$ is the length of the $m$-th step
of the primitive path. We finally obtain
the total number of configurations of one strand, confined within
a tube whose primitive path is described by the set of vectors $\{
\bm{\Delta}_m \}$. The statistical weight $W$ associated with this
state is proportional to the probability distribution:
\begin{equation}
    W(\bm{\Delta}_1,\dots,\bm{\Delta}_M) \propto
    P(\{\bm{\Delta}_m\})
    \propto e^{-\frac{1}{3}q_0 b N}
    \frac{\exp\left(-\frac{3}{2b^2 N}
    \left(\sum_{m=1}^{M}\Delta_m\right)^2\right)}
    {\left(\sum_{m=1}^{M}\Delta_m\right)^{M-1}}.\label{weight}
\end{equation}
This expression is a result corresponding to the ideal Gaussian
$P_0(\bm{R}_0)$ in equation (\ref{gauss}) for a non-entangled
phantom chain. Note that the chain end-to-end distance $\bm{R}_0$
is also the end-to-end distance of the primitive path random walk
of variable-length steps: $\sum_{m=1}^M \bm{\Delta}_m =\bm{R}_0$.
The probability distribution $P(\{\bm{\Delta}_m\})$ for a chain
confined in a reptation tube is reminiscent of a normal Gaussian,
but is in fact significantly different in the form of the exponent
and the denominator (note the modulus of the tube segment vector
$\bm{\Delta}_m$).

From equation (\ref{weight}) we obtain the formal expression for
free energy of a chain confined to a tube with the primitive path
conformation $\{\bm{\Delta}_m\}$, which implicitly depends on the
end-to-end vector $\bm{R}_0$ (the separation between crosslink
points):
 \begin{equation}
    \beta F =
    \frac{3}{2b^2 N}
    \left(\sum_{m=1}^M \Delta_m \right)^2
    +\ (M-1) \ln\left(\sum_{m=1}^{M}\Delta_m\right),
    \label{free0}
\end{equation}

\section{Free energy of deformations}

We now perform a procedure which is analogous to the one used to
obtain equation (\ref{fgauss}). In the polymer melt before
crosslinking, the ensemble of chains obeys the distribution in
(\ref{weight}) giving the free energy per strand (\ref{free0}).
The process of crosslinking not only quenches the end points of
each of the crosslinked strands, but also quenches the nodes of
the primitive path ${\bm\Delta}_m$, since the crosslinked chains
cannot disentangle due to the fixed network topology. In our mean
field approach, the tube segments described by ${\bm\Delta}_m$ are
conserved, although they may be deformed by the strains applied to
the network. For evaluating the quenched average, note that the
statistical weight (\ref{weight}) treats the tube segments $m$ in
a symmetric way. This allows us to perform the explicit summation
over the index $m$:
\begin{eqnarray}
    \beta F &=&\left\langle
     \frac{3}{2b^2 N}
    \left(\sum_{m=1}^M \Delta_m \right)^2
    + (M-1) \ln\left(\sum_{m=1}^{M}\Delta_m\right)\right\rangle\nonumber\\
    &=&\frac{3}{2b^2 N}
    \left(M \bigg\langle\Delta_m^2 \bigg\rangle
         + M(M-1) \bigg\langle\Delta_m\Delta_n \bigg\rangle \right)
    \nonumber\\
    &&+(M-1)\left\langle\ln\left(\sum_{m=1}^{M}\Delta_m\right)\right\rangle,
    \label{free}
\end{eqnarray}
where, in the second term, $m$ and $n \neq m$ are arbitrary
indices of the primitive path steps. The brackets $\langle\cdots
\rangle$ refer to the average with respect to the weight
$P(\{\bm{\Delta}_m\})$ given in (\ref{weight}).

Furthermore, note that any affine deformation $\ll$ transforms the
vectors  $\bm{\Delta}_m$ into $\bm{\Delta}_m^{'}=
\ll\bm{\Delta}_m$ and, hence, their lengths
$\Delta_m=|\bm{\Delta}_m|$ into $\Delta_m^{'}=
|\ll\bm{\Delta}_m|$, but leaves the quenched distribution $P(\{
\bm{\Delta}_m\}) $ unchanged. Bearing this in mind, we can
evaluate the averages (\ref{free}), leading to the free energy per
chain of the crosslinked network. The Appendix gives a more
detailed account of how one evaluates the averages. The resulting
average elastic energy density also incorporates the density of
crosslinked chains in the system by the constant $\mu=n_{\rm ch}k_BT$:
\begin{eqnarray}
    F_{\rm elast} &=&\frac{2}{3}\mu \,
    \frac{2M+1}{3M+1}\mathrm{Tr}(\ll^{\mathsf{T}}\ll)
    \label{free-energy}\\
    &+& \frac{3}{2}\mu \, (M-1)\frac{2M+1}{3M+1} (\overline{|\ll|})^2
    + \mu (M-1)\overline{\ln|\ll|}, \nonumber
\end{eqnarray}
where the following notations are employed:
\begin{eqnarray}
    \overline{|\ll|}&=&\frac{1}{4\pi}\int_{|\bm{e}|=1}d\Omega\
    |\ll\ \bm{e}|\label{modmat}\\
    \overline{\ln|\ll|}&=&\frac{1}{4\pi}\int_{|\bm{e}|=1}d\Omega\
    \ln|\ll\ \bm{e}|. \label{logmat}
\end{eqnarray}
The notation $\overline{\cdots}$ refers to the angular integration over a unit
vector $\bm{e}$.

In the case of uniaxial deformation, where $\ll$ takes a diagonal
form with $\lambda^{\parallel}=\lambda$ and $\lambda^{\perp}=
1/\sqrt{\lambda}$, the expressions (\ref{modmat})
and(\ref{logmat}) can be calculated explicitly. The Appendix
contains the formulae (equations (\ref{trace-eval}) -
(\ref{logmat-eval})), which need to be inserted into
(\ref{free-energy}) to give the full rubber-elastic energy
density. In the small deformation limit, where
$\lambda=1+\varepsilon$, expression (\ref{free-energy})
immediately leads to the Young's modulus $E$ in $F_{\rm
ext}\approx\frac{1}{2}E \varepsilon^2$ after noting that
\begin{eqnarray*}
    \mathrm{Tr}(\ll^{\mathsf{T}}\ll)&\approx&
        3+3\ \varepsilon^2\\
     \overline{|\ll|}&\approx&1+\frac{2}{5}\varepsilon^2\\
     \overline{\ln|\ll|}&\approx&\frac{3}{10}\varepsilon^2.
\end{eqnarray*}

For small simple shear, we can write the matrix $\ll$ as
$\lambda_{ij} =\delta_{ij}+\varepsilon u_i v_j$, where ${\bm u}$
and ${\bm v}$ are two orthogonal unit vectors. As expected, the
shear modulus $G$ in $F_{\rm shear}\approx \frac{1}{2}
G\varepsilon^2$ is one third of the Young's modulus $E$:
\begin{equation}
    G=\frac{1}{3}E=\mu\left(
        \frac{4}{3}\cdot\frac{2M+1}{3M+1}
            +\frac{1}{5}(M-1) \cdot \frac{11M+5}{3M+1}  \right) .
\label{shearmod}
\end{equation}

We point out that the expression (\ref{free-energy}) is nearly
identical to the rubber elastic free energy density obtained by
Higgs and Ball (1989), although they start from a completely
different set of physical and assumptions and mathematical
framework. In HB, the entanglements make the strands to interact
with each other only at certain points. This is modelled by
forcing the individual strands to pass through a certain number of
hoops. They effectively divide the polymer strand into segments,
each performing a three-dimensional phantom random walk. In this
approach, the entanglements only exert constraints at points and
allow the polymer to obey free Gaussian statistics between these
interaction points. Nevertheless, the apparent identity of the
final expression for the rubber-elastic free energy density is a
comforting reflection of consistency in underlying physical
concepts and the mathematical treatment of both models. The
internal similarity of the models arises from the treatment of
local chain reptation, equilibrating the distribution of its
segments between the nodes of primitive path, before and after the
deformation.

From the expression (\ref{free-energy}), we can recover the
ordinary  free energy of a phantom chain network by taking the
case $M=1$: $F_{\rm elast}=\frac{1}{2}\mu \mathrm{Tr}
(\ll^{\mathsf{T}}\ll)$. This limit means physically that the
polymer strand is placed in one single tube, tightly confined to
the axis. Mathematically, a random walk in three dimensions with
the total of $N$ steps is equivalent to a random walk in one
dimension along a given direction $\bm{\Delta}$, with $N/3$ steps,
while the two perpendicular excursions in a tightly confining
potential do not contribute to the global elastic response. This
fact is the underlying reason why we recover the phantom chain
network result by taking $M=1$ in our model. Of course, in a
parallel ``hoop model'' of HB the case $M=1$ simply means that
there are no constraints.

On the other hand, as the number of tube segments $M$ becomes very
large, we obtain a rubber-elastic elastic energy of the form
\begin{equation}
  F_{\rm elast}=
  \mu \, M\bigg( (\overline{|\ll|})^2+\overline{\ln|\ll|} \bigg).
  \label{dense-limit}
 \end{equation}
There are two ways to arrive at this limit of $M\gg 1$ in a real
physical situation: either the polymer melt is very dense, causing
a high entanglement density, or the polymer chain is very long. In
the latter case, the polymer strand experiences many confining
entanglements along its path.

Recall that the $F_{\rm elast}$ is the elastic energy density,
which relates to the free energy per chain, such as the
eq.~(\ref{free0}), and is proportional to the density of elastic
strands in the system: $F_{\rm elast} \propto \mu = n_{\rm ch}
k_BT$, where $n_{\rm ch}$ is the number of crosslinked strands per
unit volume. We can assume that in a melt, or a semidilute
solution, the chain density is inversely proportional to the
volume of an average chain, hence inversely proportional to the
chain contour length: $n_{\rm ch}\propto 1/L$. In the case of
phantom network with a free energy $F_{\rm elast}= \frac{1}{2}\mu
{\rm Tr}(\ll^{\sf T}\ll)$, one conclude that $F_{\rm elast}
\rightarrow 0$ as the chains become infinitely long! This
unphysical result reflects the fact that the phantom chain model
assumes the entanglement interactions of the chains irrelevant.
Clearly, this assumptions breaks down in the long chain limit,
where we expect the entanglements to play a crucial role.

This unphysical feature of phantom network model is overcome by
our expression (\ref{dense-limit}). As the strands become longer,
they will experience more entanglements, generating more primitive
path nodes and confining tube segments. One can assume that the
number of entanglements scales linearly with the strand length
$L$: $M\propto L$, in fact: $M = N/N_{e}$ with $N_e$ the
characteristic entanglement length, essentially the average of
$s_m^*$. Considering expression (\ref{dense-limit}), with $n_{\rm
ch}= 1/v_{\rm monomer}N$ (the inverse volume of the chain in a melt),
we note that the corresponding elastic free energy $F_{\rm elast}$
does not vanish in the limit $L \rightarrow \infty$. At $M\gg 1$
the constant plateau modulus becomes $\mu M = k_BT / v_{\rm monomer}
N_e$. As one expects, in an entangled polymer system there is no
real difference between the network and the melt and they both
have the same plateau modulus -- with the corollary that the
entanglement length $N_e$ must be much smaller in the permanently
crosslinked network.

\section{Conclusion}

In this present work, we have analysed the behaviour of a polymer
network in presence of entanglements, which are treated within a
classical mean-field tube model. We found that this leads to an
elastic energy which is identical to the one derived in the hoop
model of HB (Higgs \& Ball 1989). We claim that our model captures the
physics of entanglements in a better way than HB, since the
entanglements at a microscopic level do not so much localise the
polymer at fixed points in space along its path, but rather impede
the chain fluctuations on a length scale which is much larger than
the size of monomers. This is due to the fact that the
entanglements are caused by neighbouring strands, which likewise
fluctuate. In this sense, our model provides a firmer ground of
the theoretically known and experimentally tested results.

\begin{figure}
\centering \resizebox{0.55\textwidth}{!}{\includegraphics{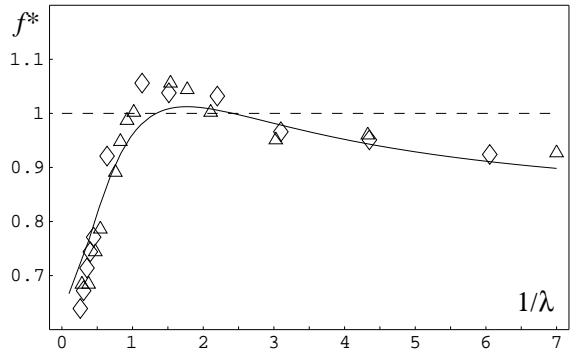}}
\caption{The Mooney-Rivlin plot: the reduced stress function
$f^*=\sigma/(\lambda-1/\lambda^2)$ plotted against $1/\lambda$.
Data points from the analysis of ref. (Gaylord \& Douglas 1990)
are fitted by the result arising from full rubber-elastic energy
(\ref{free-energy}). Fitting indicates that $M\gg 1$ and the
equation (\ref{dense-limit}) is a good approximation.}
\label{mooney}
\end{figure}

Since our results for the isotropic rubber are identical to the
ones by HB and theirs have been extensively compared with
experimental data in a review by Higgs \& Gaylord (1990), we do
not need to include a similar comparison here. We only remark that
the theoretically predicted curve for the reduced stress function
$f^*$ agrees with the experimental data to a good level, see
Fig.~\ref{mooney}.

Our model only describes the equilibrium response of a network to
deformation. Shortly after applying the deformation, traditionally
assumed affine on the level of primitive path segments, the
network finds its chains far from the microscopic equilibrium.
Each strand then redistributes its monomers between the deformed
the tube segments, attributing more monomers to some segments,
less to others. This process proceeds via the local reptation, the
sliding motion along the primitive path, and would be reflected by
a time dependence of the variable $s_{m}$. This number of monomers
attributed to the tube segment $m$ is changing, after the
deformation, between a the initial value $s_m^*$ in the
Eq.~(\ref{saddle-sm2}) and the corresponding saddle-point value at
$\Delta_m' = | \ll \bm{\Delta}_m |$. By describing the associated
relaxation process, our model can be naturally extended to
describe the short-time viscoelastic response of an ideal
crosslinked network (with no loops and dangling ends, leading to
long relaxation of stress); the characteristic time of such a
reptation dynamics is the Rouse time, since the dominant process
is the monomer diffusion along the principal path.\\

We appreciate many valuable discussions with S.~F.~Edwards.
S.~K. gratefully acknowledges support from an Overseas Research
Scholarship, from the Cambridge Overseas Trust and from Corpus
Christi College.

\appendix{}
\section*{Evaluation of quenched averages
    $\langle\Delta_{m}\Delta_{n}\rangle$ and
    $\langle\ln(\sum\Delta_{m})\rangle$}

To evaluate the thermodynamic average $\langle\Delta_m\Delta_n\rangle$,
$m,n=1,\dots,N$ in the general case, one needs to average
$\Delta_m\Delta_n=|\bm{\Delta}_m||\bm{\Delta}_n|$ with the probability
distribution given by Eq.~(\ref{weight}). For this purpose, one first has to
find the normalisation $\hat{\cal N}$ of the distribution, which
can most easily be achieved by introducing a new scalar variable
$u=\sum_{m=1}^M \Delta_m$ to simplify the exponent of this
distribution:
\begin{eqnarray*}
    \hat{\cal N}
     &=&
        \prod_{m=1}^{M}\int d\bm{\Delta}_m
            \frac{\exp\left(-\frac{3}{2b^2N}
                \left(\sum_{m=1}^{M}\Delta_m\right)^2\right)
                }{\left(\sum\Delta_m\right)^{M-1}} \nonumber\\
    &=&\prod_{m=1}^{M}\int d\bm{\Delta}_m
   \int\limits_0^{\infty} du\ \delta\left(u-\sum_{m=1}^M\Delta_m\right)
            \frac{e^{-\frac{3}{2b^2N}u^2}}{u^{M-1}}\nonumber\\
   &=&(4\pi)^{M}\int\limits_0^{\infty} du\frac{e^{-\frac{3}{2b^2N}u^2}}{u^{M-1}}
        \underline{
          \int\limits_0^u d\Delta_1 \Delta_1^2
        \int\limits_0^{u-\Delta_1}\hspace{-2mm} d\Delta_2
        \Delta_2^2\cdots}\nonumber\\
    &&  \underline{\int\limits_0^{u-\dots-\Delta_{M-2}}\hspace{-5mm}
        d\Delta_{M-1} \Delta_{M-1}^2\cdot
        (u-\Delta_1-\dots-\Delta_{M-1})^2}
\end{eqnarray*}
In the last step, we introduced spherical coordinates for the variables
$\bm{\Delta}_m$, implemented the constraint $u=\sum\Delta_m$ and used the fact
that the variables $\Delta_m$ are bound to be positive.
The underlined expression is a function of $u$, which we call $I_M(u)$.
Since the integrals only involve power functions, $I_M(u)$ itself is
a power in $u$, whose order can be determined by counting the dimensions:
\begin{displaymath}
    I_M(u)=\frac{1}{X_M}u^{3M-1}.
\end{displaymath}
After noting the recursive structure of $I_M(u)$, one can find a recursion
relation for the coefficients $X_M$:
\begin{displaymath}
    X_{M+1}=\frac{1}{2}3M(3M+1)(3M+2)X_{M}.
\end{displaymath}
Since $I_{M=1}(u)=u^2$, all coefficients $X_M$ and hence all
functions $I_M(u)$ are known.
The remaining integration of $u$ is a standard Gaussian integral.

Having obtained the normalisation constant $\hat{\cal N}$, we can proceed to
calculate the averages $\langle\Delta_{m}\Delta_{n}\rangle$. The calculations
are very similar to the above one for $\hat{\cal N}$; the only significant
difference is the
angular part of the integration which produces terms of the form
$\frac{1}{3} \mathrm{Tr}(\ll^{\mathsf{T}}\ll)$ and $\overline{|\ll|}$ for the
case $m=n$ or $m\not=n$ respectively.

For the logarithmic term, one in fact finds that only the angular
integration yields relevant terms:
\begin{eqnarray*}
 \lefteqn{\left\langle\ln\left(\sum_{m=1}^M\Delta_m\right)\right\rangle=}
     \nonumber\\
    && \frac{1}{\hat{\cal N}}
       \int_0^{\infty}du\frac{e^{-\frac{3}{2b^2N}u^2}}{u^{M-1}}
        \int_0^u d\Delta_1 \Delta_1^2
        \int_0^{u-\Delta_1}\hspace{-5mm} d\Delta_2
        \Delta_2^2\cdots\nonumber\\
    &&  \int_0^{u-\dots-\Delta_{M-2}}\hspace{-15mm}
        d\Delta_{M-1} \Delta_{M-1}^2\
        (u-\Delta_1-\dots-\Delta_{M-1})^2\\
    && \left(\prod_{m=1}^{M}\int d\Omega_m\right)
        \ln[
        \Delta_1|\ll\bm{e}_1|+\dots+\Delta_{M-1}|\ll\bm{e}_{M-1}|
        \nonumber\\
    && \hspace{28mm}+(u-\Delta_1-\dots-\Delta_{M-1})|\ll \bm{e}_M|]
\end{eqnarray*}
with the unit vector $\bm{e}_m$ specifies the (arbitrary)
orientation of the corresponding tube segment $\bm{\Delta}_m$.
Then one observes:
\begin{eqnarray*}
\lefteqn{ \ln \left[ \Delta_1|\ll\bm{e}_1| +\dots
+\Delta_{M-1}|\ll\bm{e}_{M-1}|
+(u-\Delta_1-\dots-\Delta_{M-1})|\ll \bm{e}_M| \right] }
  \nonumber\\
    &=&\ln[u]+\ln[|\ll\bm{e}_M|]+
\ln\Bigg[1+\underbrace{\sum_{m=1}^{M-1}
        \left(\frac{|\ll\bm{e}_m|}{|\ll\bm{e}_M|}-1\right)
            \frac{\Delta_m}{u}}_{\mbox{small since }u\gg|\bm{\Delta}_m|}\Bigg]
            \nonumber\\
&\approx&\ln[|\ll\bm{e}_M|]+\mathrm{const}.
\end{eqnarray*}

In case of uniaxial deformation, the evaluation of $\mathrm{Tr}
(\ll^{\mathsf{T}}\ll)$, $\overline{|\ll|}$ and $\overline{\ln|\ll|}$,
cf. equations (\ref{free-energy}) -- (\ref{logmat}) has
been given elsewhere (Higgs \& Gaylord 1990), but for completeness
we state them here as well:
\begin{eqnarray}
    \mathrm{Tr}(\ll^{\mathsf{T}}\ll)
    &=& \lambda^2+\frac{1}{\lambda}   \label{trace-eval}\\
    \overline{|\ll|}
    &=&\frac{1}{2}\Bigg(\lambda
        +\frac{1}{2\sqrt{\lambda}\sqrt{\lambda^3-1}}
        \ln\left(\frac{\lambda^{3/2}+\sqrt{\lambda^3-1}}
        {\lambda^{3/2}-\sqrt{\lambda^3-1}}\right) \Bigg)  \label{modmat-eval}\\
    \overline{\ln(|\ll|)}&=&
    \ln(\lambda)-1+\frac{\arctan \sqrt{\lambda^3-1}}{\sqrt{\lambda^3-1}},
     \label{logmat-eval}
\end{eqnarray}

\end{document}